\documentclass[preprint,12pt]{elsarticle}

\usepackage{amssymb}
\usepackage{multirow}

\begin{document}

\begin{frontmatter}



\title{Contemporary Recommendation Systems on Big Data and Their Applications: A Survey}


\author[label1]{Ziyuan Xia}

\affiliation[label1]{organization={Antai College of Economics and Management, Shanghai Jiao Tong University},
            addressline={1954 Huashan Road}, 
            city={Shanghai},
            postcode={200030}, 
            state={Shanghai},
            country={China}}
            
\author[label2]{Anchen Sun}
 
\affiliation[label2]{organization={Department of Electrical and Computer Engineering, University of Miami},
 	addressline={1251 Memorial Drive}, 
 	city={Coral Gables},
 	postcode={33146}, 
 	state={FL},
 	country={USA}}
 	
\author[label3]{Jingyi Xu}

\affiliation[label3]{organization={Department of Architecture, Cornell University},
	addressline={129 Sibley Dome},
	city={Ithaca},
	postcode={14853}, 
	state={NY},
	country={USA}}

\author[label2]{Yuanzhe Peng}

\author[label4]{Rui Ma}

\affiliation[label4]{organization={Bascom Palmer Eye Institute, University of Miami Miller School of Medicine},
	addressline={900 NW 17th St}, 
	city={Miami},
	postcode={33136}, 
	state={FL},
	country={USA}}

\author[label5,label6]{Minghui Cheng}

\affiliation[label5]{organization={Department of Civil \& Architectural Engineering, University of Miami},
	addressline={1251 Memorial Drive}, 
	city={Coral Gables},
	postcode={33146}, 
	state={FL},
	country={USA}}

\affiliation[label6]{organization={Department of Architecture, University of Miami},
	addressline={1223 Theo Dickinson Dr}, 
	city={Coral Gables},
	postcode={33146}, 
	state={FL},
	country={USA}}
	
\begin{abstract}
This survey paper conducts a comprehensive analysis of the evolution and contemporary landscape of recommendation systems, which have been extensively incorporated across a myriad of web applications. It delves into the progression of personalized recommendation methodologies tailored for online products or services, organizing the array of recommendation techniques into four main categories: content-based, collaborative filtering, knowledge-based, and hybrid approaches, each designed to cater to specific contexts. The document provides an in-depth review of both the historical underpinnings and the cutting-edge innovations in the domain of recommendation systems, with a special focus on implementations leveraging big data analytics. The paper also highlights the utilization of prominent datasets such as MovieLens, Amazon Reviews, Netflix Prize, Last.fm, and Yelp in evaluating recommendation algorithms. It further outlines and explores the predominant challenges encountered in the current generation of recommendation systems, including issues related to data sparsity, scalability, and the imperative for diversified recommendation outputs. The survey underscores these challenges as promising directions for subsequent research endeavors within the discipline. Additionally, the paper examines various real-life applications driven by recommendation systems, addressing the hurdles involved in seamlessly integrating these systems into everyday life. Ultimately, the survey underscores how the advancements in recommendation systems, propelled by big data technologies, have the potential to significantly enhance real-world experiences.
\end{abstract}

\begin{graphicalabstract}
\end{graphicalabstract}

\begin{highlights}
\item Comprehensive Analysis and Methodological Classification: It provides a thorough exploration of the evolution and current landscape of recommendation systems, categorizing the techniques into content-based, collaborative filtering, knowledge-based, and hybrid approaches. This classification is designed to address the unique requirements of various online products and services, emphasizing the role of big data analytics in driving innovations within this domain.

\item Evaluation Using Prominent Datasets and Challenges: The paper emphasizes the use of key datasets, such as MovieLens, Amazon Reviews, Netflix Prize, Last.fm, and Yelp, for evaluating the performance of recommendation algorithms. It also outlines significant challenges faced by modern recommendation systems, including data sparsity, scalability issues, and the need for diversified recommendations, positioning these challenges as avenues for future research.

\item Real-life Applications and Impact on Society: By examining the integration of recommendation systems in various domains such as marketing, governance, and healthcare, the paper underscores their significance in enhancing real-world experiences. It highlights how advancements in recommendation technologies, propelled by big data, are shaping user experiences and influencing societal trends, offering insights into their potential for future societal impact.
\end{highlights}


\end{frontmatter}


\section{Introduction}
\label{sec:introduction}
\begin{figure}[htbp]
	\centering
	\includegraphics[width=0.8\textwidth, trim=100 10 100 10,clip]{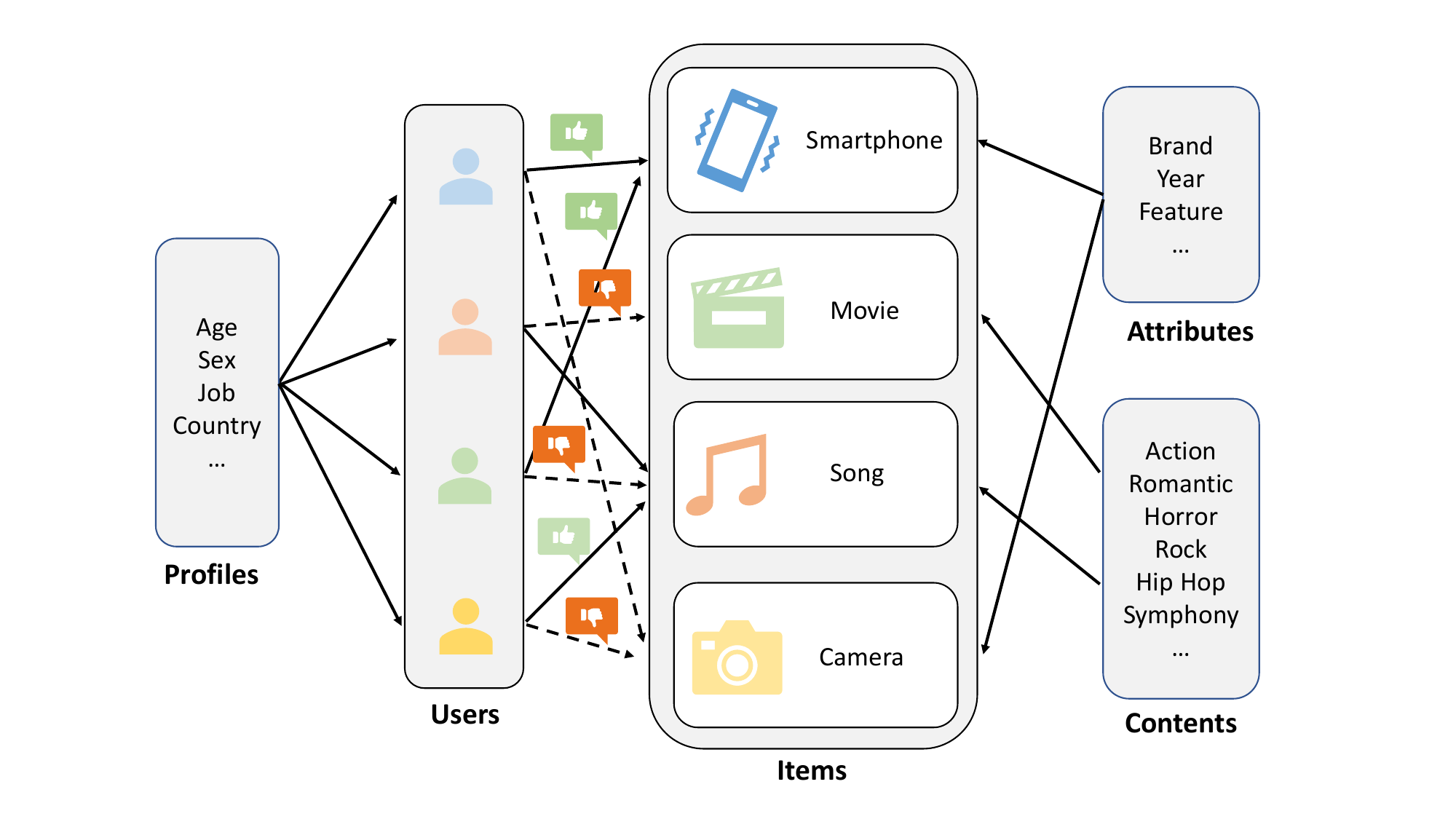}
	\caption{The logical process of a modern recommendation system.}
	\label{fig:fig1}
\end{figure}

In this survey, we scrutinize the escalating popularity and diverse applications of recommendation systems in web applications, a topic extensively explored by Zhou et al. \cite{zhou2021combinatorial}. These systems, a specialized category of information filtering systems, are designed to predict user preferences for various items. They play a pivotal role in guiding decision-making processes, such as purchasing decisions and music selections, as discussed by Wang et al. \cite{wang2020personalized}. An exemplary instance of this application is Amazon's personalized recommendation engine, which tailors each user's homepage. Major corporations like Amazon, YouTube, and Netflix utilize these systems to enhance user experience and generate substantial revenue, as highlighted by Adomavicius et al. and Omura et al. \cite{adomavicius2005toward, omura2020ad}. Figure \ref{fig:fig1} from Entezari et al. \cite{entezari2021tensor} illustrates a modern recommendation system. Additionally, these systems are increasingly pertinent in the field of human-computer interaction (HCI), where they improve interaction efficiency through feedback mechanisms, a topic explored in several studies \cite{7962152}.

Recommendation systems are particularly crucial for certain companies, as their efficiency can lead to substantial revenue generation and competitive advantage, as evidenced in the research by Rismanto et al. and Cui et al. \cite{rismanto2020research, cui2020personalized}. For example, Netflix's \textquotedblleft{Netflix Prize}\textquotedblright{} challenge aimed to develop a recommender system surpassing their existing algorithm, offering a substantial prize to incentivize innovation.

Furthermore, in the domain of big data, recommendation systems are highly prevalent, as detailed by Li et al. \cite{li2021research, li2021sports}. These systems predict user interests in purchasing based on extensive data analysis, including purchase history, ratings, and reviews. There are four widely recognized types of recommendation systems, as identified by Numnonda \cite{numnonda2018real}: content-based, collaborative filtering-based, knowledge-based, and hybrid-based, each with distinct advantages and drawbacks, as Xiao et al. elucidate \cite{xiao2018personalized}. For example, collaborative filtering-based systems may face issues such as data sparsity and scalability, as Huang et al. mention \cite{huang2019multimodal}, and cold-start problems, while content-based systems might struggle to diversify user interests, as noted by Zhang et al.~\cite{zhang2018mcrs} and Benouaret et al.~\cite{benouaret2020comparative}.

As recommendation system algorithms, neural networks, and big data continue to evolve, their integration into various aspects of our lives is increasingly evident. We are witnessing a surge in applications that leverage big data, recommendation systems, and machine learning to enhance our daily experiences. In the realm of marketing, these technologies enable more personalized recommendations that align with individuals' personal backgrounds and cultural preferences, whether they are watching movies, shopping online, or traveling. The impact extends to governance, where big data applications are facilitating more efficient and responsive public services. Simultaneously, these innovations are promoting a more sustainable lifestyle and fostering stronger connections in healthcare, demonstrating their potential to improve well-being across various sectors. Moreover, these applications are collecting vast amounts of data across all age groups, from children to seniors, empowering individuals with insights to better understand and analyze their behaviors and preferences~\cite{sun2024said, huh2019location}. This data-driven approach is revolutionizing how we interact with technology, making it a pivotal element in our journey towards a more informed and connected society.

The structure of this paper is outlined as follows: Section \ref{Recommendation Systems} provides a comprehensive exploration of recommendation systems, tracing their evolution from historical foundations to contemporary state-of-the-art methodologies, accompanied by a thorough examination of recent advancements within the sector. Section \ref{Recommendation System based on Big Data} addresses the specific challenges encountered in recommendation systems that leverage big data, such as data sparsity, scalability issues, and the imperative for diverse recommendations. This section also delves into potential solutions for overcoming these obstacles. Section \ref{Recommendation System Application} is dedicated to the application of recommendation systems in real-life contexts, discussing their practical implications and the challenges associated with their integration. Finally, the paper culminates with a comprehensive summary presented in Section \ref{Summary}.

\section{Recommendation Systems}
\label{Recommendation Systems}

Recommendation systems aim to predict users' preferences for a certain item and provide personalized services \cite{yi2019deep}. This section will discuss several commonly used recommender methods, such as content-based method, collaborative filtering-based method, knowledge-based method, and hybrid-based method.

\subsection{Content-based Recommendation Systems}
\label{Content-based Recommendation Systems}

The main idea of content-based recommenders is to recommend items based on the similarity between different users or items \cite{lops2011content}. This algorithm determines and differentiates the main common attributes of a particular user's favorite items by analyzing the descriptions of those items. Then, these preferences are stored in this user's profile. The algorithm then recommends items with a higher degree of similarity with the user's profile. Besides, content-based recommendation systems can capture the specific interests of the user and can recommend rare items that are of little interest to other users. However, since the feature representations of items are designed manually to a certain extent, this method requires a lot of domain knowledge. In addition, content-based recommendation systems can only recommend based on users' existing interests, so the ability to expand users' existing interests is limited.

\begin{figure}[htbp]
	\centering
	\includegraphics[width=0.7 \textwidth, trim=140 80 140 80,clip]{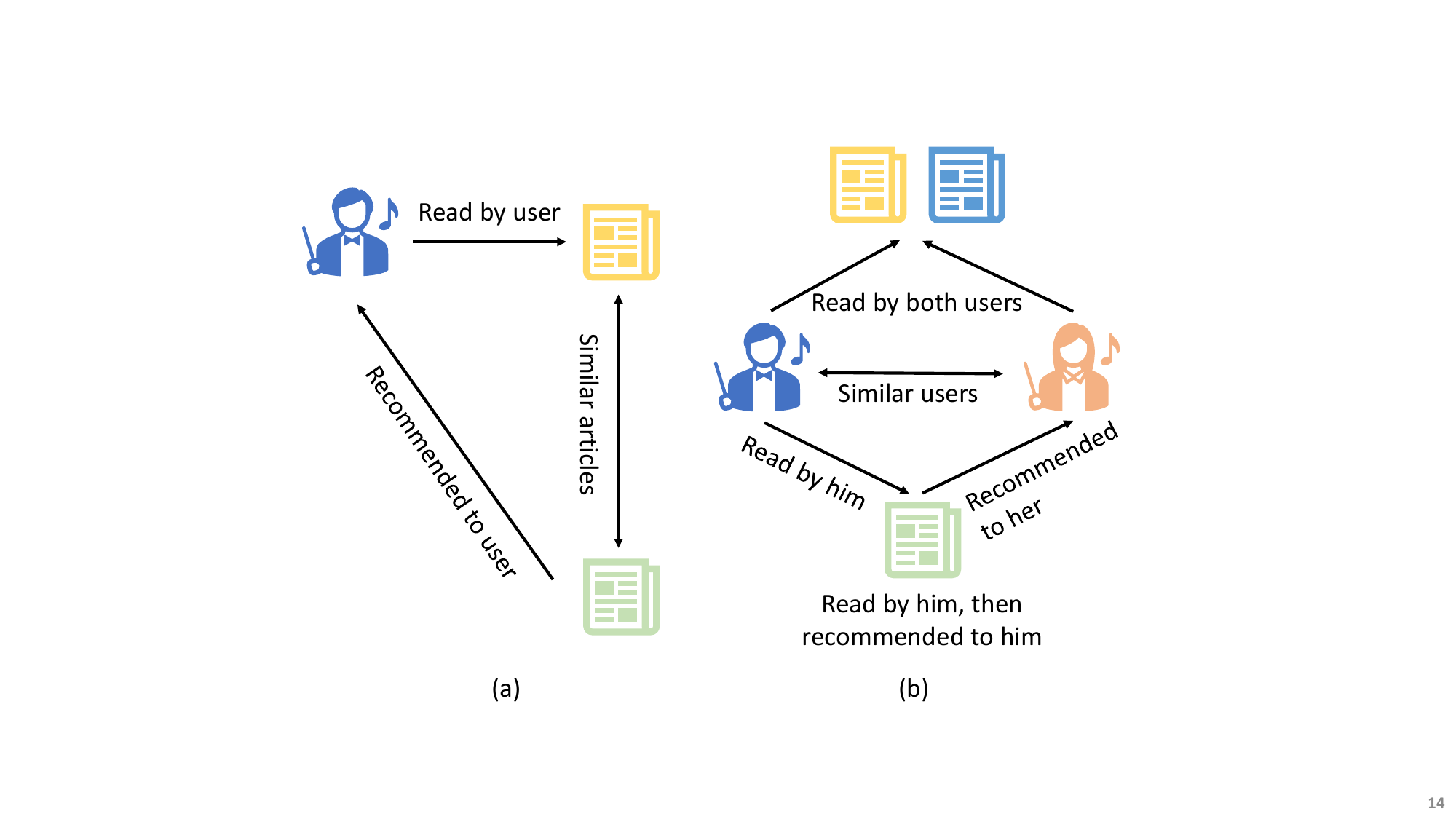}
	\caption{(a) Content-based Recommendation System (b) Collaborative Filtering-based Recommendation System.}
	\label{fig2}
\end{figure}

\subsection{Collaborative Filtering-based Recommendation Systems}

Collaborative Filtering-based (CF) methods are primarily used in big data processing platforms due to their parallelization characteristics \cite{elahi2016survey}. The basic principle of the recommendation system based on collaborative filtering is shown in Fig. \ref{fig2} \cite{alhijawi2016using}. CF recommendation systems use the behavior of a group of users to recommend to other users \cite{al2018similarity}. There are mainly two types of collaborative filtering techniques, which are user-based and item-based.

\begin{itemize}
	\item User-based CF: In the user-based CF recommendation system, users will receive recommendations of products that similar users like \cite{rezaimehr2021survey}. Many similarity metrics can calculate the similarity between users or items, such as Constrained Pearson Correlation coefficient (CPC), cosine similarity, adjusted cosine similarity, etc. For example, cosine similarity is a measure of similarity between two vectors. Let $x$ and $y$ denote two vectors, cosine similarity between $x$ and $y$ can be represented by
	
	\begin{equation}
		\cos{\left(\theta\right)}=\frac{\sum_{i=1}^{n}{x_iy_i}}{\sqrt{\sum_{i=1}^{n}x_i^2}\sqrt{\sum_{i=1}^{n}y_i^2}}
	\end{equation}
	
	\begin{equation}
		r_{xy}=\frac{\sum_{i=1}^{n}{{(x}_i-\bar{x})}{(y}_i-\bar{y})}{\sqrt{\sum_{i=1}^{n}{{(x}_i-\bar{x})}^2}\sqrt{\sum_{i=1}^{n}{{(y}_i-\bar{y})}^2}}
	\end{equation}
	
	\item Item-based CF: Item-based CF algorithm predicts user ratings for items based on item similarity. Generally, item-based CF yields better results than user-based CF because user-based CF suffers from sparsity and scalability issues. However, both user-based CF and item-based CF may suffer from cold-start problems \cite{zhang2016fast}.
	
\end{itemize}

\begin{table}[]
	\centering
	\caption{Summary of the Modern Recommendation Systems Methods.}
	\begin{tabular}{|l|l|l|}
		\hline
		\begin{tabular}[c]{@{}l@{}}Recommendation \\ Systems\end{tabular}           & Descriptive Key Points                                                                                                                            & Papers                                                                            \\ \hline
		Content-based                 & \begin{tabular}[c]{@{}l@{}}Recommend items based on\\ the similarity between\\ different items.\end{tabular}                                      & \begin{tabular}[c]{@{}l@{}} Musto et al. \cite{musto2016learning}\\ Volkovs et al.\cite{volkovs2017content}\\ Mittal et al.\cite{mittal2020smart}\\ Almaguer et al.\cite{perez2021content}\end{tabular} \\ \hline
		\begin{tabular}[c]{@{}l@{}}Collaborative \\ Filtering-based\end{tabular} & \begin{tabular}[c]{@{}l@{}}Recommend items to some\\ users based on the other\\ users behavior.\end{tabular}                                      & \begin{tabular}[c]{@{}l@{}} Zhang et al.\cite{zhang2019privacy}\\  Bobadilla et al.\cite{bobadilla2020deep}\\  Bobadilla et al.\cite{bobadilla2020classification}\\ Rezaimehr et al.\cite{rezaimehr2021survey} \end{tabular}         \\ \hline
		Knowledge-based               & \begin{tabular}[c]{@{}l@{}}Recommend items to users\\ based on basic knowledge of\\ users, items, and relationships\\ between items.\end{tabular} & \begin{tabular}[c]{@{}l@{}} Dong et al.\cite{dong2020interactive}\\  Gazdar et al.\cite{gazdar2020new}\\  Alamdari et al.\cite{alamdari2020systematic}\\ Cena et al.\cite{cena2021logical} \end{tabular}         \\ \hline
		Hybrid-based                  & \begin{tabular}[c]{@{}l@{}}Recommend items to users\\ based on more than one\\ filtering approach.\end{tabular}                                   & \begin{tabular}[c]{@{}l@{}} Hrnjica et al.\cite{hrnjica2020model}\\  Shokeen et al.\cite{shokeen2020study}\\  Zagra et al.\cite{zagranovskaia2021designing}\\ Ibrahim et al.\cite{ibrahim2021hybrid} \end{tabular}         \\ \hline
	\end{tabular}
\end{table}

\subsection{Knowledge-based Recommendation Systems}
\label{Knowledge-based Recommendation Systems}

The main idea of knowledge-based recommendation systems is to recommend items to users based on basic knowledge of users, items, and relationships between items \cite{shishehchi2012ontological, aggarwal2016knowledge}. Since knowledge-based recommendation systems do not require user ratings or purchase history, there is no cold start problem for this type of recommendation \cite{cabezas2017knowledge}. Knowledge-based recommendation systems are well suited for complex domains where items are not frequently purchased, such as cars and apartments \cite{tarus2018knowledge}. But the acquisition of required domain knowledge can become a bottleneck for this recommendation technique \cite{dong2020interactive}.

\subsection{Hybrid-based Recommendation Systems}

Hybrid-based recommendation systems combine the advantages of multiple recommendation techniques and aim to overcome the potential weaknesses in traditional recommendation systems \cite{ribeiro2012pareto}. There are seven basic hybrid recommendation techniques \cite{ibrahim2021hybrid}: weighted, mixed, switching, feature combination, feature augmentation, cascade, and meta-level methods \cite{hassan2016enhancing, zhang2016hybrid}. Among all of these methods, the most commonly used is the combination of the CF recommendation methods with other recommendation methods (such as content-based or knowledge-based) to avoid sparsity, scalability, and cold-start problems \cite{hrnjica2020model, zagranovskaia2021designing, george2019review}.

\subsection{Challenges in Modern Recommendation Systems}
\label{Challenges in Modern Recommendation Systems}

\begin{itemize}
	\item Sparsity. As we know, the usage of recommendation systems is growing rapidly. Many commercial recommendation systems use large datasets, and the user-item matrix used for filtering may be very large and sparse. Therefore, the performance of the recommendation process may be degraded due to the cold start problems caused by data sparsity \cite{west2016recommendation}.
	\item Scalability. Traditional algorithms will face scalability issues as the number of users and items increases. Assuming there are millions of customers and millions of items, the algorithm's complexity will be too large. However, recommendation systems must respond to the user's needs immediately, regardless of the user's rating history and purchase situation, which requires high scalability. For example, Twitter is a large web company that uses clusters of machines to scale recommendations for its millions of users \cite{shokeen2020study}.
	\item Diversity. Recommendation systems also need to increase diversity to help users discover new items. Unfortunately, some traditional algorithms may accidentally do the opposite because they always recommend popular and highly-rated items that some specific users love. Therefore, new hybrid methods need to be developed to improve the performance of the recommendation systems \cite{he2021research}.
\end{itemize}

\section{Recommendation System based on Big Data}
\label{Recommendation System based on Big Data}

\begin{figure}[htbp]
	\centering
	\includegraphics[width=0.6\textwidth, trim=200 130 250 130,clip]{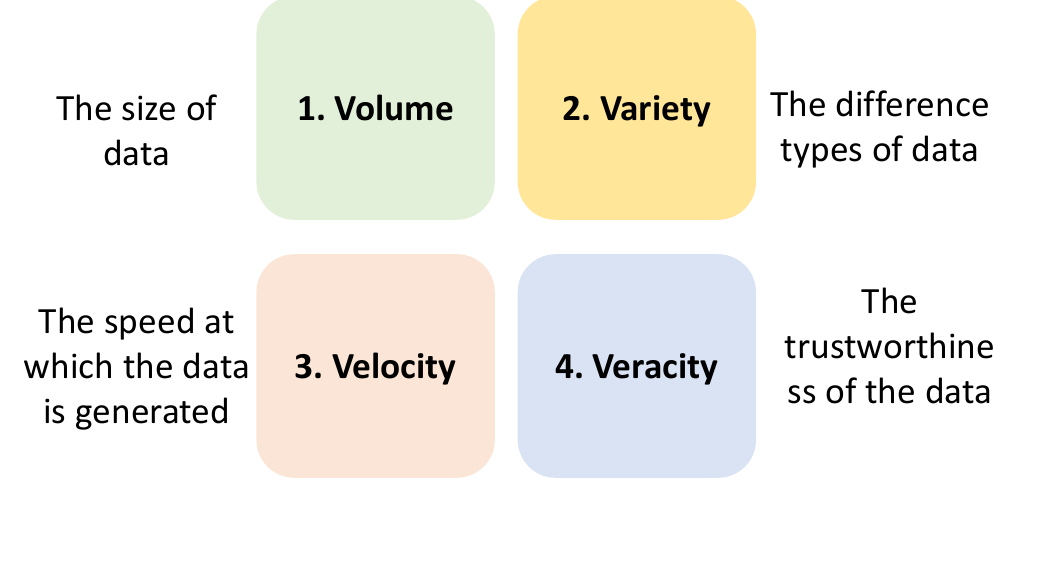}
	\caption{The 4V of big data.}
	\label{fig3}
\end{figure}

Big data refers to the massive, high growth rate and diversified information \cite{chen2021dqn, kadam2016big}. It requires new processing models to have stronger decision-making and process optimization capabilities \cite{acharjya2016survey}. Big data has its unique “4V” characteristics, as shown in Fig. \ref{fig3} \cite{zhou2018academic}: Volume, Variety, Velocity, and Veracity.

\subsection{Big Data Processing Flow}
\begin{figure}[htbp]
	\centering
	\includegraphics[width=0.8\textwidth, trim=180 40 200 30,clip]{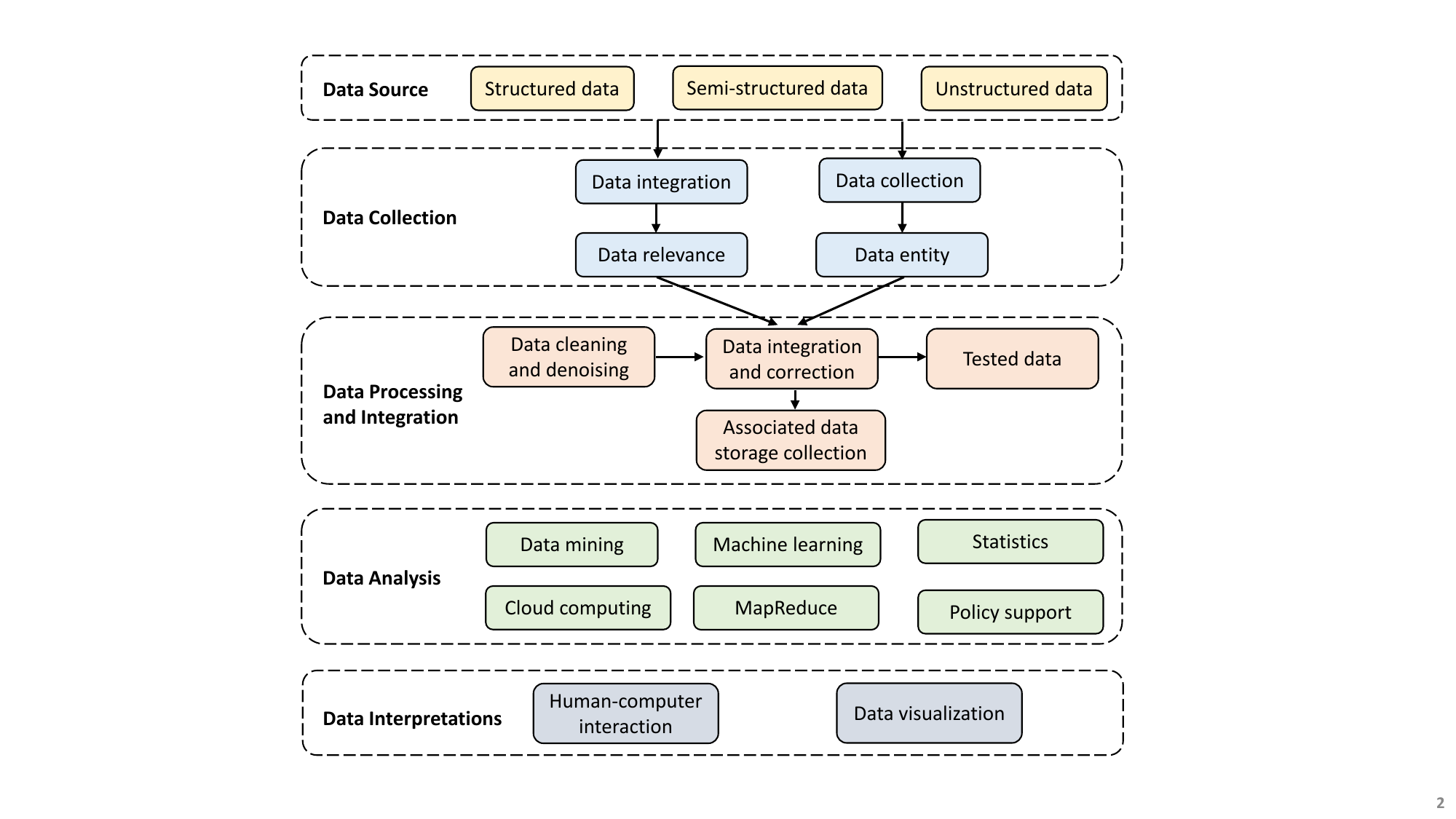}
	\caption{The basic flow of big data processing.}
	\label{fig4}
\end{figure}

Big data comes from many sources, and there are many methods to process it \cite{ram2018privacy}. However, the primary processing of big data can be divided into four steps \cite{wu2013data}. Besides, Fig. \ref{fig4} presents the basic flow of big data processing.

\begin{itemize}
	\item Data Collection.
	\item Data Processing and Integration. The collection terminal itself already has a data repository, but it cannot accurately analyze the data. The received information needs to be pre-processed \cite{emani2015understandable}.
	\item Data Analysis. In this process, these initial data are always deeply analyzed using cloud computing technology \cite{lin2019cloud}.
	\item Data Interpretation.
\end{itemize}

\subsection{Modern Recommendation Systems based on the Big Data}

The shortcomings of traditional recommendation systems mainly focus on insufficient scalability and parallelism \cite{cheng2020research}. For small-scale recommendation tasks, a single desktop computer is sufficient for data mining goals, and many techniques are designed for this type of problems \cite{al2021hybrid}.

\begin{figure}[htbp]
	\centering
	\includegraphics[width=0.95 \textwidth, trim=30 50 10 50,clip]{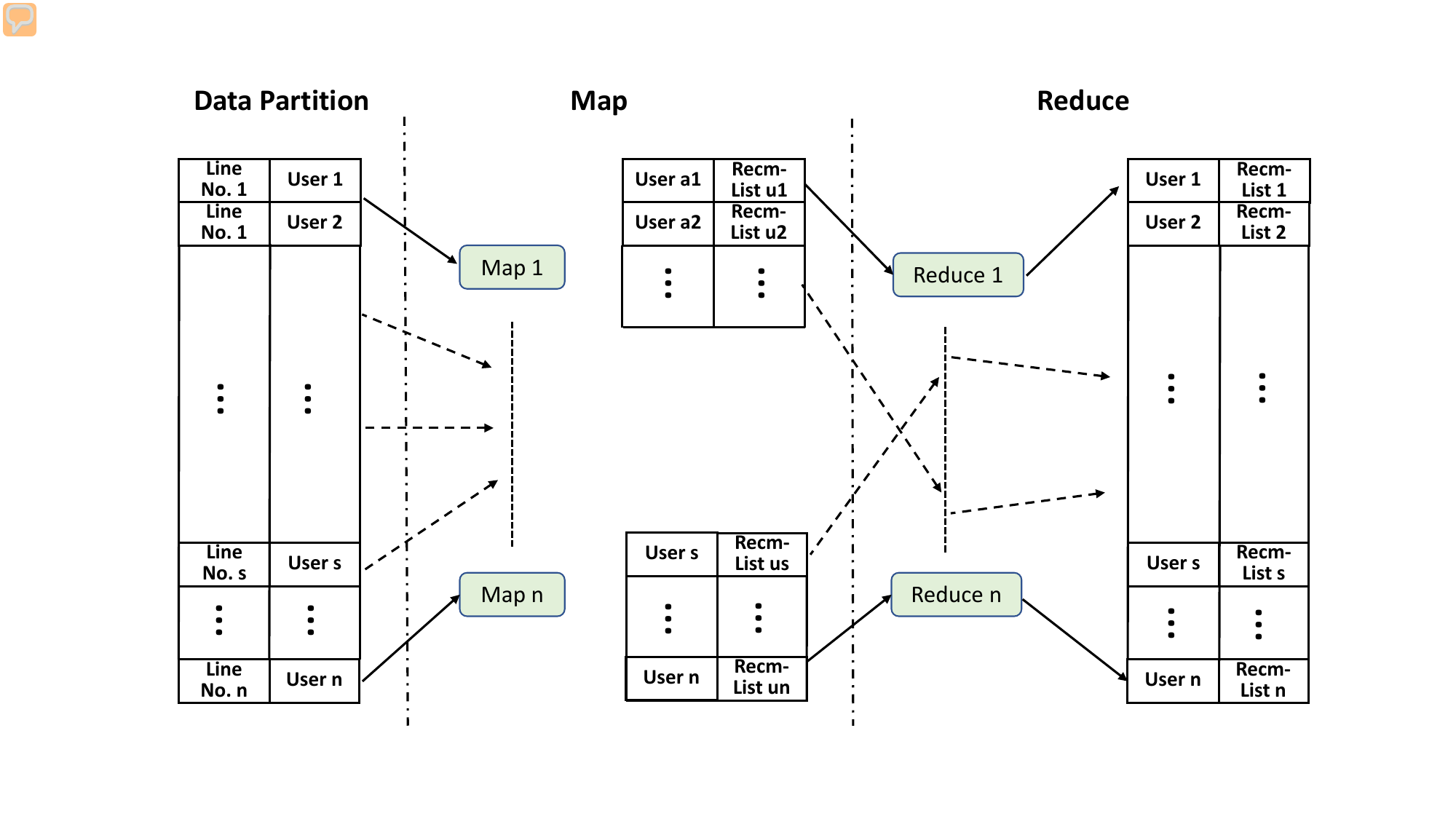}
	\caption{Mapreduce in the Recommendation Systems.}
	\label{fig5}
\end{figure}

However, the rating data is usually so large for medium-scale recommendation systems that it is impossible to load all the data into memory at once \cite{dev2016recommendation}. Common solutions are based on parallel computing or collective mining, sampling and aggregating data from different sources, and using parallel computing programming to perform the mining process \cite{chen2018disease}. The big data processing framework will rely on cluster computers with high-performance computing platforms \cite{wan2021research}. At the same time, data mining tasks will be deployed on a large number of computing nodes (i.e., clusters) by running some parallel programming tools \cite{asiyakeyword}, such as MapReduce \cite{kadam2016big, verma2015big}. For example, Fig. \ref{fig5} is the MapReduce in the Recommendation Systems.

\begin{figure}[htbp]
	\centering
	\includegraphics[width=0.85\textwidth, trim=10 10 10 10,clip]{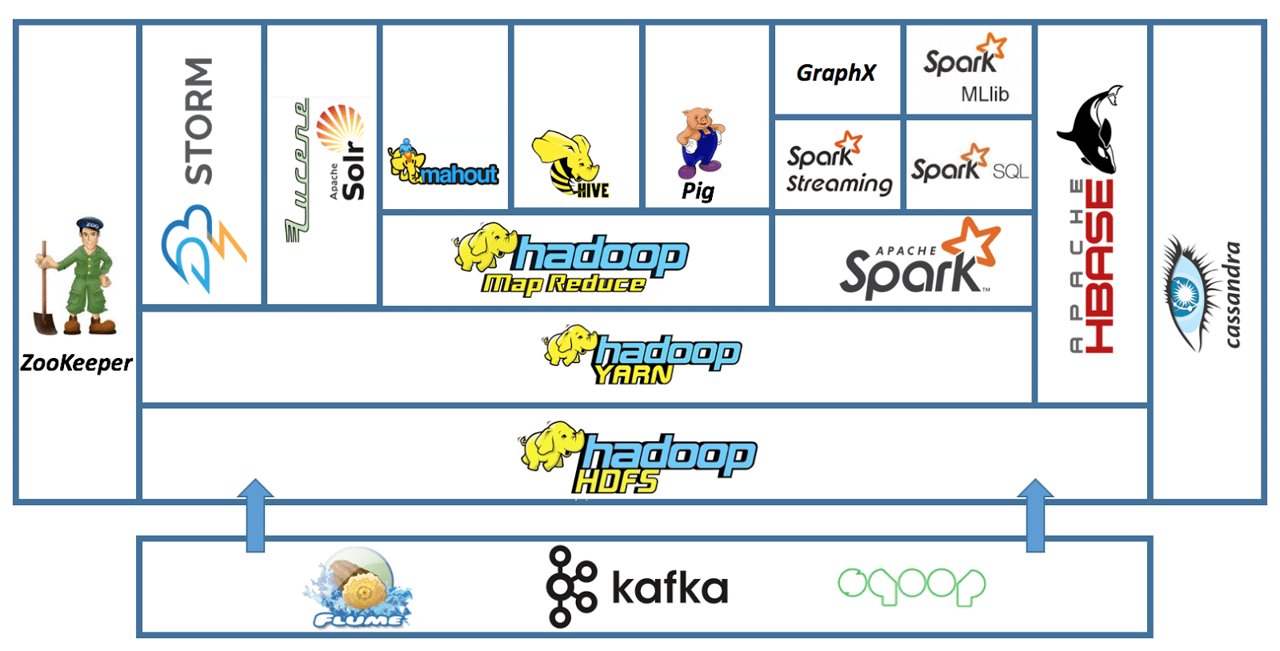}
	\caption{The ecosystem of Apach Hadoop.}
	\label{fig6}
\end{figure}

In recent years, various big data platforms have emerged \cite{uzun2021big}. For example, Hadoop and Spark \cite{kadam2016big}, both developed by the Apache Software Foundation, are widely used open-source frameworks for big data architectures \cite{kadam2016big, zhang2018service}. Each framework contains an extensive ecosystem of open-source technologies that prepare, process, manage and analyze big data sets \cite{chaithra2015user}. For example, Fig. \ref{fig6} is the ecosystem of Apache Hadoop \cite{ait2019distributed}.

Hadoop allows users to manage big data sets by enabling a network of computers (or \textquotedblleft{nodes}\textquotedblright) to solve vast and intricate data problems. It is a highly scalable, cost-effective solution that stores and processes structured, semi-structured and unstructured data.

Spark is a data processing engine for big data sets. Like Hadoop, Spark splits up large tasks across different nodes. However, it tends to perform faster than Hadoop, and it uses random access memory (RAM) to cache and process data instead of a file system. This enables Spark to handle use cases that Hadoop cannot. The following are some benefits of the Spark framework:

\begin{itemize}
	\item It is a unified engine that supports SQL queries, streaming data, machine learning (ML), and graph processing.
	\item It can be 100x faster than Hadoop for smaller workloads via in-memory processing, disk data storage, etc.
	\item It has APIs designed for ease of use when manipulating semi-structured data and transforming data.
\end{itemize}

\begin{figure}[htbp]
	\centering
	\includegraphics[width=0.95\textwidth, trim=130 160 10 160,clip]{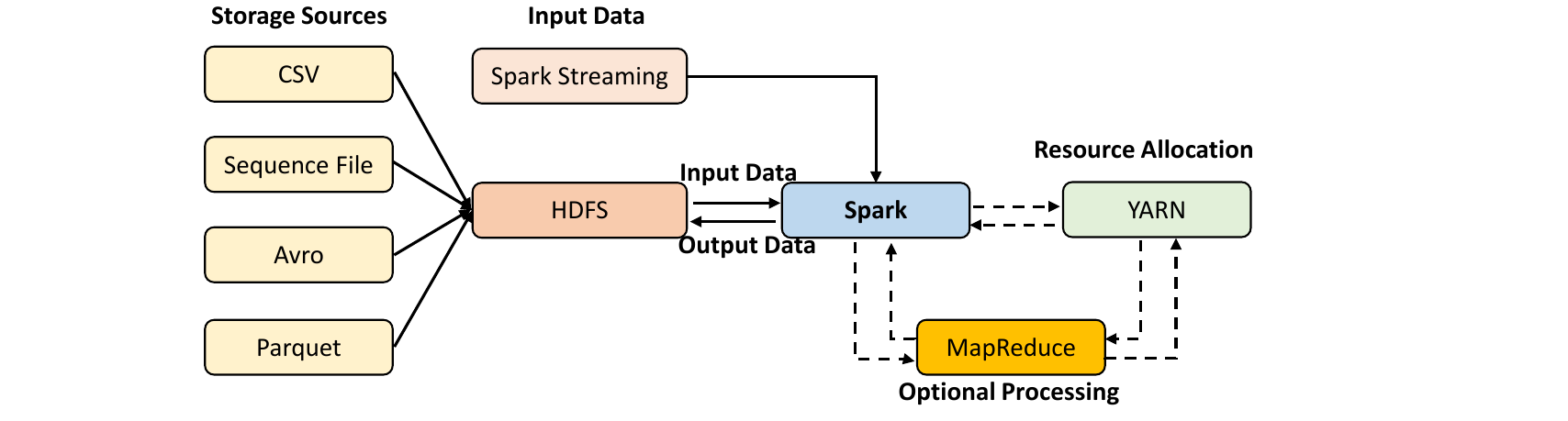}
	\caption{Spark uses the best parts of Hadoop through HDFS for reading and storing data, MapReduce for optional processing and YARN for resource allocation.}
	\label{fig7}
\end{figure}

\begin{figure}[htbp]
	\centering
	\includegraphics[width=0.8 \textwidth, trim=200 60 200 40,clip]{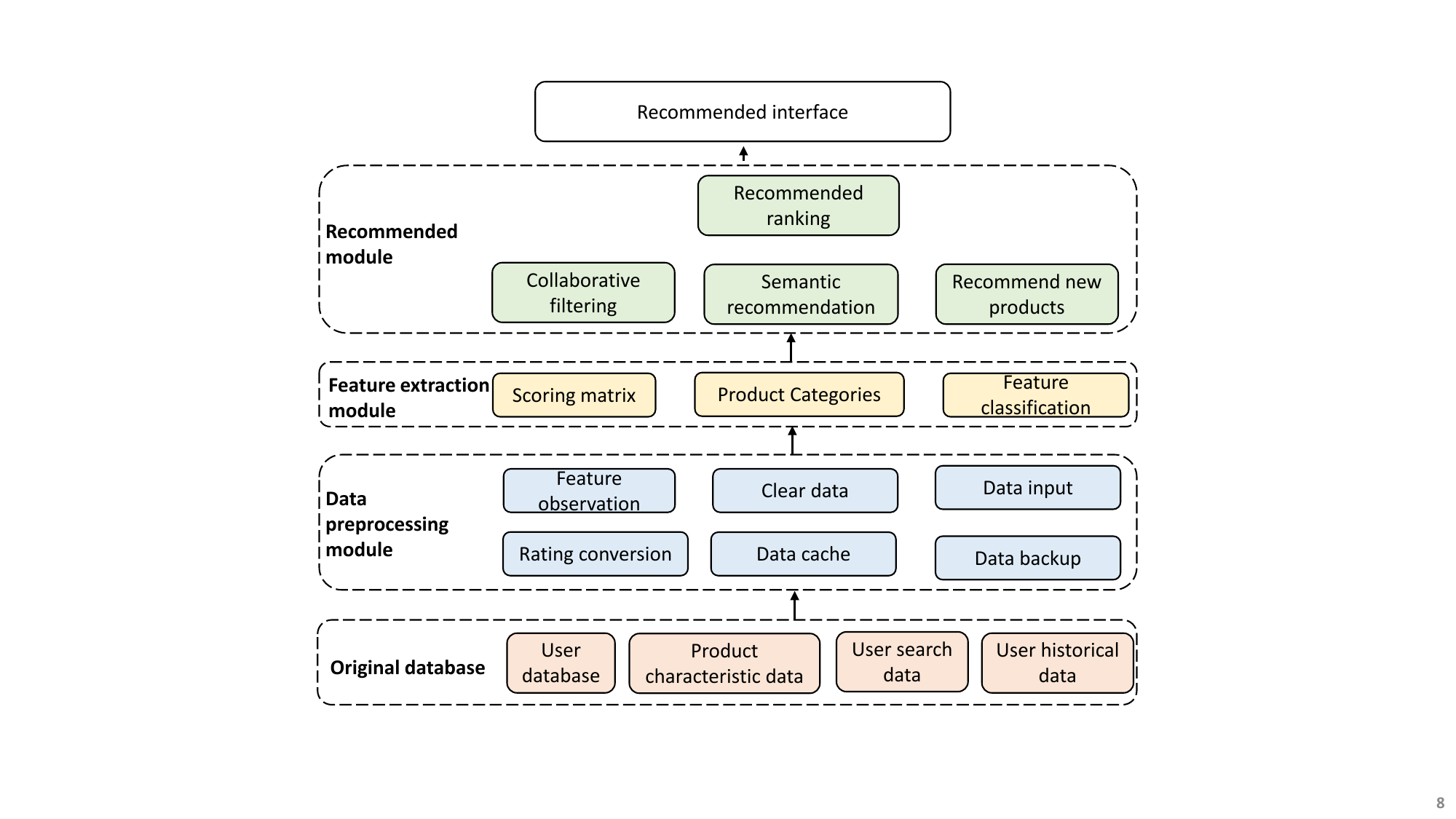}
	\caption{The architecture of the modern recommendation system based on Spark.}
	\label{fig8}
\end{figure}

Furthermore, Spark is fully compatible with the Hadoop eco-system and works smoothly with Hadoop Distributed File System (HDFS), Apache Hive, and others. Thus, when the data size is too big for Spark to handle in-memory, Hadoop can help overcome that hurdle via its HDFS functionality. Fig. \ref{fig7} is a visual example of how Spark and Hadoop can work together. Fig. \ref{fig8} is the the architecture of the modern recommendation system based on Spark.

\section{Applications of Modern Recommendation System}
\label{Recommendation System Application}

Recommendation systems have become ubiquitous across diverse sectors such as search engines, digital media platforms, and e-commerce sites on the internet [1][2]. The progression of information technology has led to their significant evolution, embracing increasingly complex models. The emergence of big data has been a catalyst for the refinement of recommendation systems, empowering them to offer more precise and holistic recommendations. Currently, the incorporation of big data into contemporary recommendation systems plays a crucial role in streamlining operations in e-commerce and e-governance, as well as in fostering sustainable living practices [3][4][5]. This integration signifies a leap forward in leveraging technology to enhance decision-making processes and user experiences across various digital platforms.

\subsection{Recommendation System in E-Commerce}

Recommendation systems have evolved from specialized tools utilized by a select few e-commerce platforms to vital commercial assets that significantly transform the e-commerce landscape [6]. Major online platforms and applications, such as Amazon and TikTok, now harness the power of big data to refine their recommendation algorithms for users [7][8]. Ansari et al. previously highlighted the need for advancements in data collection and analytics to expand the operational advantages in the marketing sector, prompted by the advent of new applications for information agents [9]. With the advancement of big data, the application of information agents has shifted towards providing accurate and tailored recommendations to customers on online markets and platforms through innovative models like topic modeling and sentiment analysis [10][11]. Therefore, in an era dominated by big data, it's evident that recommender systems have become widely integrated into various aspects of e-commerce operations. This extensive adoption emphasizes the pivotal role these systems play in improving user experience, personalizing customer interactions, and boosting the overall efficiency of e-commerce platforms.

\subsection{Recommendation System in E-Governance}

E-governance stands as a fundamental challenge in the realm of smart city initiatives, integrating information technology and big data in the public sector to elevate the delivery of services and information. This approach not only aims to enhance government transparency, accountability, and trustworthiness but also to engage citizens in the governance process [12]. In the digital era, particularly highlighted by the rise of epidemics, the demand for e-governance in society is on the rise. This underscores the necessity for governments to establish adept online information systems to meet the goals of effective e-governance [13]. Businesses, including pharmaceutical companies, are also recognizing the need for digital governance. For instance, they can implement recommendation systems, built on blockchain and machine learning technologies, to streamline drug shipment monitoring [14]. Additionally, these systems have applications in demand-side management, like energy management, where they utilize big data analytics to identify residential users' preferences for energy-efficient appliances [15]. Thus, the deployment of recommendation systems, bolstered by big data technologies, plays a crucial role in improving the digital governance framework, optimizing business processes, and facilitating efficient energy management in the digital age.

\subsection{Recommendation System in Sustainable Lifestyle}

In the context of an academic discourse exploring the impact of recommendation systems on fostering a sustainable lifestyle, the digital transformation catalyzed by the Internet revolution has significantly facilitated the transition towards e-commerce and a lifestyle centered around digital interactions, concurrently with a heightened emphasis on sustainability across environment, societal, and governance sectors~\cite{xia2023dynamic}. This shift has intensified the demand for environmentally friendly practices in daily activities, such as online shopping, dietary habits, and transportation, prompting both organizations and individuals to prefer products and services with reduced environmental footprints. Addressing the nexus of technological innovation, consumer behavior, and environmental responsibility, recent studies advocate the integration of green marketing strategies with online retail platforms through the deployment of sophisticated recommendation systems~\cite{felfernig2023recommender}. These systems are instrumental in refining the digital shopping journey, aiming to promote sustainable living by prioritizing options with lower environmental impacts, thus facilitating a shift towards sustainability among both providers and consumers. The work of Lee et al~\cite{lee2011recommender}. and Zhang et al~\cite{zhanga2022towards}. highlights the critical capacity of recommendation systems to advocate for environmentally sustainable choices, representing a crucial advancement in embedding eco-sustainability within the digital commerce ecosystem.

Furthermore, green building emerges as a significant aspect that profoundly influences our connection to sustainable living~\cite{xu2021sustainable}. With the advancement of recommendation systems, big data, and the Internet of Things (IoT), several challenges associated with green building can be addressed through the integration of recommendation systems and machine learning technologies. These technologies hold the potential to enhance various facets of green building, including:

\begin{itemize}
	\item Enhancing energy efficiency within buildings,~\cite{himeur2021survey, siddique2023smartbuild}
	\item Facilitating the selection and acquisition of green buildings,~\cite{gogoi2018green}
	\item Implementing Building Information Modeling (BIM) and Life Cycle Assessment (LCA) for sustainable construction,~\cite{veselka2020recommendations, cheng2023data}
	\item Innovating in the realm of green building design.~\cite{domingo2023recommender}
	
\end{itemize}

Such integrations not only underscore the potential of technological innovations in promoting energy efficiency and supporting sustainable building practices but also mark a significant stride towards leveraging recommendation systems in the pursuit of a greener lifestyle.

\subsection{Recommendation System in Healthcare}
The healthcare industry is experiencing a paradigm shift with the incorporation of cutting-edge technologies, and recommendation systems have emerged as a pivotal component in this transformative landscape. Leveraging machine learning algorithms and data analytics, recommendation systems in healthcare offer a versatile range of applications, from improving clinical decision-making to enhancing patient engagement. This subsection explores the diverse facets of recommendation systems in healthcare and their impact on various stakeholders within the ecosystem.

\subsubsection{Clinical Decision Support}.
One of the primary applications of recommendation systems in healthcare is in clinical decision support. These systems analyze electronic health records (EHRs), medical literature, and patient data to assist healthcare professionals in making informed decisions about diagnostics, treatment plans, and interventions. By providing relevant and evidence-based information, recommendation systems contribute to more accurate and personalized patient care, potentially reducing diagnostic errors and improving overall healthcare outcomes~\cite{bates2003impact, shojania2009effects}.

\subsubsection{Patient-Centered Care}.
In the era of patient-centered care, recommendation systems play a crucial role in tailoring healthcare services to individual patient needs. These systems analyze patient preferences, demographics, and health histories to generate personalized recommendations for treatment options, preventive measures, and lifestyle modifications. By fostering patient engagement and empowerment, recommendation systems contribute to a more collaborative and effective healthcare relationship between providers and patients~\cite{coulter2007effectiveness, street2009communication}.

\subsubsection{Resource Optimization}.
Recommendation systems help optimize healthcare resources by streamlining processes and improving operational efficiency. For instance, in hospital management, these systems can suggest optimal bed allocation, appointment scheduling, and resource utilization based on historical data and real-time information. By minimizing bottlenecks and enhancing resource allocation, recommendation systems contribute to cost-effectiveness and improved service delivery~\cite{li2017survey, brailsford2009comprehensive}.

\subsubsection{Telehealth and Remote Monitoring}.
With the rise of telehealth and remote monitoring, recommendation systems support healthcare providers in delivering virtual care. These systems analyze patient-generated health data from wearable devices, monitoring tools, and telehealth platforms to provide timely recommendations for interventions, medication adjustments, or lifestyle modifications. This real-time support contributes to proactive healthcare management, particularly for patients with chronic conditions~\cite{fatehi2018clinician, dorsey2016state}.

\subsubsection{Collaborative Healthcare Networks}.
Recommendation systems facilitate collaboration and knowledge sharing among healthcare professionals through the creation of collaborative networks. By analyzing expertise, research interests, and clinical experiences, these systems connect healthcare professionals for consultations, research collaborations, and second opinions. This fosters a culture of continuous learning and knowledge dissemination within the healthcare community~\cite{alshammary2019collaborative, choi2018systematic}.

As recommendation systems continue to evolve, their integration into healthcare processes holds immense potential for improving patient outcomes, enhancing operational efficiency, and advancing the overall quality of healthcare services. However, challenges such as data interoperability, privacy concerns, and algorithmic transparency must be addressed to ensure the responsible and ethical deployment of recommendation systems in the complex healthcare landscape. Ongoing research, interdisciplinary collaboration, and stakeholder engagement are crucial for harnessing the full benefits of recommendation systems in healthcare.


\section{Datasets for Recommendation Systems}
Datasets play a pivotal role in the development, evaluation, and benchmarking of recommendation systems. The choice of dataset depends on factors such as the application domain, the type of recommendation task, and the specific research objectives. In this section, we provide a summary of some commonly used datasets, as shown in Table.~\ref{tab:datasets}, in recommendation systems research:

\begin{table}[]
\begin{tabular}{|l|l|l|l|l|}
\hline
Dataset                         & Top Models                                                               & Metrics           & \begin{tabular}[c]{@{}l@{}}Extra \\ Training \\ Data\end{tabular} & Year \\ \hline
\multirow{3}{*}{MovieLens 100K} & GHRS~\cite{GHRS}                                                                     & 0.887(RSME)       & N                                                                 & 2021 \\ \cline{2-5} 
                                & GLocal-K~\cite{GlocalK}                                                                 & 0.889(RSME)       & N                                                                 & 2021 \\ \cline{2-5} 
                                & MG-GAT~\cite{MGGAT}                                                                   & 0.890(RSME)       & Y                                                                 & 2020 \\ \hline
\multirow{3}{*}{MovieLens 1M}   & GLocal-K~\cite{GlocalK}                                                                 & 0.823(RSME)       & N                                                                 & 2021 \\ \cline{2-5} 
                                & Sparse FC~\cite{SparseFC}                                                                & 0.824(RSME)       & N                                                                 & 2018 \\ \cline{2-5} 
                                & CF-NADE~\cite{CFNADE}                                                                  & 0.829(RSME)       & N                                                                 & 2016 \\ \hline
\multirow{3}{*}{MovieLens 10M}  & \begin{tabular}[c]{@{}l@{}}Bayesian \\ timeSVD++ \\ flipped\end{tabular}~\cite{BayesiantimeSVD++} & 0.749(RSME)       & N                                                                 & 2019 \\ \cline{2-5} 
                                & \begin{tabular}[c]{@{}l@{}}Bayesian \\ timeSVD++\end{tabular}~\cite{BayesiantimeSVD++}            & 0.752(RSME)       & N                                                                 & 2019 \\ \cline{2-5} 
                                & \begin{tabular}[c]{@{}l@{}}Bayesian \\ SVD++\end{tabular}~\cite{BayesiantimeSVD++}                & 0.756(RSME)       & N                                                                 & 2019 \\ \hline
\multirow{3}{*}{MovieLens 20M}  & VASP~\cite{VASP}                                                                     & 0.448(nDCG@100)   & Y                                                                 & 2021 \\ \cline{2-5} 
                                & \begin{tabular}[c]{@{}l@{}}H+Vamp\\ Gated\end{tabular}~\cite{H+VAMP}                   & 0.445(nDCG@100)   & Y                                                                 & 2019 \\ \cline{2-5} 
                                & RecVAE~\cite{Recvae}                                                                   & 0.442(nDCG@100)   & N                                                                 & 2019 \\ \hline
\multirow{3}{*}{Amazon-Book}    & SSCF~\cite{SSCF}                                                                     & 0.065(nDCG@20)    & N                                                                 & 2022 \\ \cline{2-5} 
                                & SANSA~\cite{SANSA}                                                                    & 0.064(nDCG@20)    & N                                                                 & 2023 \\ \cline{2-5} 
                                & BSPM-LM~\cite{BSPMLM}                                                                  & 0.061(nDCG@20)    & N                                                                 & 2022 \\ \hline
\multirow{3}{*}{Netflix}        & \begin{tabular}[c]{@{}l@{}}H+Vamp\\ Gated\end{tabular}~\cite{H+VAMP}                   & 0.409(nDCG@100)   & N                                                                 & 2019 \\ \cline{2-5} 
                                & RecVAE~\cite{Recvae}                                                                   & 0.394(nDCG@100)   & N                                                                 & 2019 \\ \cline{2-5} 
                                & EASE~\cite{EASE}                                                                     & 0.393(nDCG@100)   & N                                                                 & 2019 \\ \hline
\multirow{3}{*}{Last.FM}        & Ekar~\cite{EKAR}                                                                     & 0.248(HR@10)      & N                                                                 & 2019 \\ \cline{2-5} 
                                & HAKG~\cite{HAKG}                                                                     & 0.093(nDCG@20)    & N                                                                 & 2022 \\ \cline{2-5} 
                                & KGNN-LS~\cite{KGNNLS}                                                                  & 0.370(Recall@100) & N                                                                 & 2019 \\ \hline
Yelp                            & DGRec~\cite{DGRec}                                                                    & 0.1427(nDCG)      & N                                                                 & 2019 \\ \hline
\multirow{3}{*}{Gowalla}        & BSPM-EM~\cite{BSPM-EM}                                                                  & 0.160(nDCG@20)    & N                                                                 & 2022 \\ \cline{2-5} 
                                & MGDCF~\cite{Mgdcf}                                                                    & 0.159(nDCG@20)    & N                                                                 & 2024 \\ \cline{2-5} 
                                & UltraGCN~\cite{UltraGCN}                                                                 & 0.158(nDCG@20)    & N                                                                 & 2021 \\ \hline
\end{tabular}
\caption{Datasets and compared methods for recommendation systems}
\label{tab:datasets}
\end{table}

\subsection{MovieLens}
The MovieLens dataset is one of the most widely used datasets for collaborative filtering-based recommendation systems. It contains movie ratings provided by users, along with movie metadata such as titles, genres, and release years \cite{movielens}. MovieLens datasets are available in various sizes (e.g. 100K, 1M, 10M, 20M) ranging from small-scale datasets suitable for initial experimentation to large-scale datasets for comprehensive evaluations.

\subsection{Amazon Reviews}
Amazon provides extensive datasets comprising user reviews and ratings for a wide range of products available on its platform, including books, electronics, clothing, and more \cite{amazon_reviews}. These datasets offer rich information about user preferences, product attributes, and user-item interactions, making them valuable resources for research in recommendation systems, sentiment analysis, and e-commerce analytics.

\subsection{Netflix Prize Dataset}
The Netflix Prize dataset is a large-scale dataset of movie ratings collected from Netflix users. It was released as part of the Netflix Prize competition, a renowned competition aimed at improving the performance of recommendation algorithms \cite{netflix_prize}. The dataset contains millions of ratings provided by users over several years, along with additional contextual information such as user demographics and temporal dynamics.

\subsection{Last.fm Dataset}
The Last.fm dataset comprises music listening histories of users on the Last.fm platform, including details about artists, albums, and user preferences \cite{lastfm}. It offers insights into user behavior and preferences in the context of music consumption, making it valuable for research in music recommendation systems, personalized playlists, and music discovery applications.

\subsection{Yelp Dataset}

The Yelp dataset consists of user reviews and ratings for businesses across various categories, including restaurants, hotels, and local services \cite{yelp}. It provides rich information about user opinions, business attributes, and geographic locations, enabling research in recommendation systems, sentiment analysis, and location-based services.

\subsection{Other Datasets}

In addition to the datasets mentioned above, several other datasets are available for recommendation systems research, covering diverse domains such as news articles, social media interactions, academic papers, and online retail transactions. These datasets offer valuable opportunities for exploring different recommendation scenarios, addressing various challenges, and advancing the state-of-the-art in recommendation technology.

In summary, the availability of diverse datasets facilitates the development and evaluation of recommendation systems across different application domains. Researchers can leverage these datasets to build robust, scalable, and personalized recommendation algorithms, ultimately enhancing user experiences and driving innovation in the field of recommendation systems.

\section{Open Questions and Future Research Directions}
Despite the significant advancements made in recommendation systems on big data, several open questions persist, and numerous avenues for future research beckon. This section delves into some of these intriguing areas, presenting potential directions for researchers to explore further.

\subsection{Interpretable Recommender Systems}
Research could focus on developing recommendation models that not only provide accurate predictions but also offer transparent explanations for their recommendations~\cite{explanation_methods}. Techniques such as model-agnostic explanation methods, rule-based systems, or attention mechanisms can be explored to enhance interpretability.

\subsection{Fairness and Bias}
Future research might concentrate on devising fairness-aware recommendation algorithms that mitigate bias and ensure equitable treatment across diverse user groups~\cite{fairness_recsys}. This could involve developing fairness metrics, fairness-aware loss functions, or debiasing techniques tailored for recommendation systems.

\subsection{Context-Aware Recommendations}
Advanced research may explore leveraging deep learning architectures, such as recurrent neural networks (RNNs) or transformers, to effectively capture and utilize contextual information in recommendation models~\cite{context_aware_recsys}. Additionally, techniques like multi-modal learning could be investigated to incorporate various types of context, such as textual, visual, or temporal cues.

\subsection{Cold-Start Problem}
Novel research directions could involve exploring transfer learning approaches, where knowledge learned from related domains or auxiliary data sources is transferred to alleviate the cold-start problem~\cite{transfer_learning_recsys}. Hybrid recommendation techniques combining collaborative filtering, content-based methods, and knowledge graphs could also be investigated to handle cold-start scenarios more effectively.

\subsection{Long-Tail Recommendations}
Research efforts could focus on developing specialized algorithms for long-tail recommendations, such as mixture models, probabilistic graphical models, or ensemble methods tailored to capture rare item preferences~\cite{long_tail_recsys}. Techniques like active learning or diversity-promoting algorithms could also be explored to enhance the coverage of long-tail items in recommendations.

\subsection{Dynamic and Adaptive Recommendations}
Future research might investigate reinforcement learning frameworks for recommendation systems, where agents learn optimal recommendation policies through interaction with users and feedback~\cite{reinforcement_learning_recsys}. Additionally, online learning algorithms capable of continuously updating recommendation models in response to evolving user preferences and feedback could be explored.

\subsection{Privacy-Preserving Recommendations}
Research could explore techniques such as federated learning, where recommendation models are trained across decentralized data sources without sharing raw user data~\cite{privacy_preserving_recsys}. Differential privacy mechanisms could also be integrated into recommendation algorithms to ensure individual user privacy while maintaining recommendation utility.

\subsection{Multi-Stakeholder Recommendations}
Advanced research may focus on developing multi-objective optimization frameworks for recommendation systems, considering the diverse interests of users, providers, and advertisers simultaneously~\cite{multi_stakeholder_recsys}. Game-theoretic approaches could be employed to model interactions among stakeholders and design recommendation strategies that optimize collective utility while addressing conflicting objectives.

In conclusion, the field of recommendation systems on big data is ripe with opportunities for future exploration and innovation. By addressing these open questions and pursuing novel research directions, researchers can propel the field forward, advancing the state-of-the-art in recommendation technology and unlocking new possibilities for personalized user experiences in diverse application domains.

\section{Summary}
\label{Summary}

This survey paper offers an exhaustive overview of recommendation systems, a technological innovation that has seen widespread adoption in various web-based applications in recent years. The primary objective of modern recommendation systems is to furnish users with personalized suggestions for online products or services, employing a range of techniques including content-based, collaborative filtering, knowledge-based, and hybrid approaches to cater to diverse requirements across different scenarios.

The manuscript delves into a comprehensive historical review and a critical examination of the state-of-the-art methodologies in recommendation systems, with a special emphasis on the pioneering developments brought about by the advent of big data analytics. Notably, this paper highlights the utilization of prominent datasets such as MovieLens, Amazon Reviews, Netflix Prize, Last.fm, and Yelp in evaluating recommendation algorithms. Additionally, this paper scrutinizes the prevalent challenges encountered in contemporary recommendation systems, such as data sparsity, scalability issues, and the need for diversity in recommendations, proposing these hurdles as fertile grounds for future research endeavors.

Simultaneously, this survey extends its analysis to the application of recommendation systems within various life-related domains, including marketing, governance, medical, health, and the promotion of sustainable lifestyles. This exploration aims to provide a foundational understanding of how recommendation systems intersect with everyday life, highlighting their significance in shaping user experiences and influencing societal trends. Through this holistic review, the paper endeavors to present a nuanced perspective on the evolution of recommendation systems and their growing impact on digital consumer culture and beyond.



\bibliographystyle{elsarticle-num} 
\bibliography{main}


\end{document}